\documentclass[twocolumn,pra,superscriptaddress]{revtex4-1}

\usepackage{graphicx}
\usepackage{dcolumn}
\usepackage{bm}

\usepackage{epsf}

\usepackage{mathtools}

\usepackage{amsfonts}
\usepackage{algorithmic}

\usepackage{textcomp}

\usepackage{float}
\usepackage{textcomp}
\usepackage{setspace}
\usepackage{psfrag}
\usepackage{color,soul}
\usepackage{algorithm}
\usepackage{bbold}
\usepackage{verbatim}
\usepackage{amsthm}
\usepackage{braket}
\usepackage{enumitem}

\theoremstyle{definition}

\newcommand{\abs}[1]{\left\vert#1\right\vert}

\begin{document}

\preprint{APS/123-QED}

\title{Multiplicative Bell Inequalities}

\author{Amit Te'eni}
\author{Bar Y. Peled}%
\author{Avishy Carmi}
\affiliation{Center for Quantum Information Science and Technology \& Faculty
of Engineering Sciences, Ben-Gurion University of the Negev, Beersheba
8410501, Israel}
\author{Eliahu Cohen}
\affiliation{Faculty of Engineering and the Institute of Nanotechnology and Advanced
Materials, Bar Ilan University, Ramat Gan 5290002, Israel}
\affiliation{
Physics Department, Centre for Research in Photonics, University
of Ottawa, Advanced Research Complex, 25 Templeton, Ottawa, K1N 6N5,
Canada}

\date{\today}

\begin{abstract}

Bell inequalities are important tools in contrasting classical and quantum
behaviors. To date, most Bell inequalities are linear combinations
of statistical correlations between remote parties. Nevertheless,
finding the classical and quantum mechanical (Tsirelson) bounds for a
given Bell inequality in a general scenario is a difficult task which rarely leads
to closed-form solutions.
Here we introduce a new class of Bell inequalities based on products of correlators that alleviate
these issues. Each such Bell inequality is associated with
a unique coordination game. In the simplest case,
Alice and Bob, each having two random variables, attempt to maximize the area of a rectangle and the
rectangle's area is represented by a certain parameter. This parameter, which is a
function of the correlations between their random variables, is shown to be a Bell parameter, i.e. the achievable bound using only classical correlations is strictly smaller than the achievable bound
using non-local quantum correlations 
We continue by generalizing to the case in which Alice and Bob, each
having now $n$ random variables, wish to maximize a certain volume in
$n$-dimensional space. We term this parameter a multiplicative Bell
parameter and prove its Tsirelson bound. 
Finally, we investigate the case of
local hidden variables and show that for any deterministic strategy of
one of the players the Bell parameter is a harmonic function whose
maximum approaches the Tsirelson bound as the number of measurement
devices increases. Some theoretical and experimental implications of these results are discussed.
%
\end{abstract}

\pacs{Valid PACS appear here}
\maketitle


\section{Introduction}
Bell inequalities \cite{Bellnonlocality:14} are mathematical
instruments, which enable to find out whether correlations between
distant experimenters are stronger than those allowed by local hidden
variable theories. In other words, a violation of some Bell inequality
implies that the observed system exhibits a quantum behavior. Since
their first appearance in Bell's paper \cite{BellEPR}, Bell
inequalities have revolutionized our understanding of quantum
nonlocality in particular, and quantum theory in general. Many
variations and generalizations of Bell's original inequality have
appeared \cite{peres1999all}. 

Related research avenues in the foundations of quantum mechanics have been the search for bounds on the strength of quantum correlations \cite{buhrman2005causality,wehner2006tsirelson,navascues2008convergent,bancal2012framework,goh2018geometry,de2015simple}, as well as finding some deeper physical reasons for these bounds \cite{PR1,linden2007quantum,pawlowski2009information,navascues2009glance,oppenheim2010uncertainty,fritz2013local,van2013implausible,carmi2018relativistic}. Tsirelson bounds \cite{Tsirelson} set the maximal possible values for Bell parameters in quantum mechanics, i.e. they tell us to which extent a Bell's inequality can be violated by a quantum mechanical system.

In \cite{carmi2018significance} a novel approach has been employed for
constructing the Bell-CHSH parameter and deriving a richer Tsirelson
bound for it, which depends on local uncertainty relations. Their
prescription is as follows: one can begin by writing down a certain
covariance matrix (encoding generalized uncertainty relations),
continue by assuming that it is positive-semidefinite, and then use
the sum of quadratic forms in order to infer the inequality.

Geometrically, as any covariance may be visualized as an ellipsoid in the Euclidean space, it follows that all Bell inequalities which are linear combinations of correlations (refered to here as \textit{additive} Bell inequalities), are the weighted sum of the ellipsoid axes. This insight gave us the idea to explore another type of Bell inequalities - \textit{multiplicative} Bell inequlaities, which correspond to the ellipsoid's volume. It thus seems only natural that a better understanding of the geometry of nonlocal correlations requires both additive and multiplicative variants to be developed. As far as we know, this paper presents the first account of such inequalities. It is noteworthy that some previous papers, such as~\cite{pozsgay2017covariance}, present nonlinear Bell inequalities; however, their Bell parameters are linear combinations of covariances, and thus they are nonlinear in the one-point correlators, yet linear with respect to the two-point correlators.

Thus, in the present work we construct Bell inequalities by utilizing a product of quadratic forms rather than a sum. The Bell parameter obtained in this procedure has a Tsirelson bound which can be readily found. We are also able to conceive a game, which describes
a specific computational task that is equivalent to maximizing a
certain parameter. In more detail, Alice and Bob are engaged in a
two-player coordination game, where their decisions control the
movement of a walker over a two-dimensional grid. The objective is to
maximize the average area of the rectangle covered by the walker,
which is proportional to the value of our Bell parameter. If in this
game Alice and Bob share an EPR state, they may cover an average area
double in size compared to the outcome of a classical strategy.

This procedure can be naturally generalized to construct Bell
parameters which describe the case where Alice and Bob each have
multiple measurement devices. These parameters correspond to
appropriately generalized games, where the objective is to maximize
the average volume of a hyperrectangle. Remarkably, as the
dimensionality increases, classical and quantum strategies become
asymptotically similar, with the ratio of corresponding parameters
being at least $\sqrt{\pi/2e}$. 

\section{\label{sec:level2}The volume maximization game}
\begin{figure*}[htb]
	\centering
	
	\includegraphics[width=0.57\textwidth]{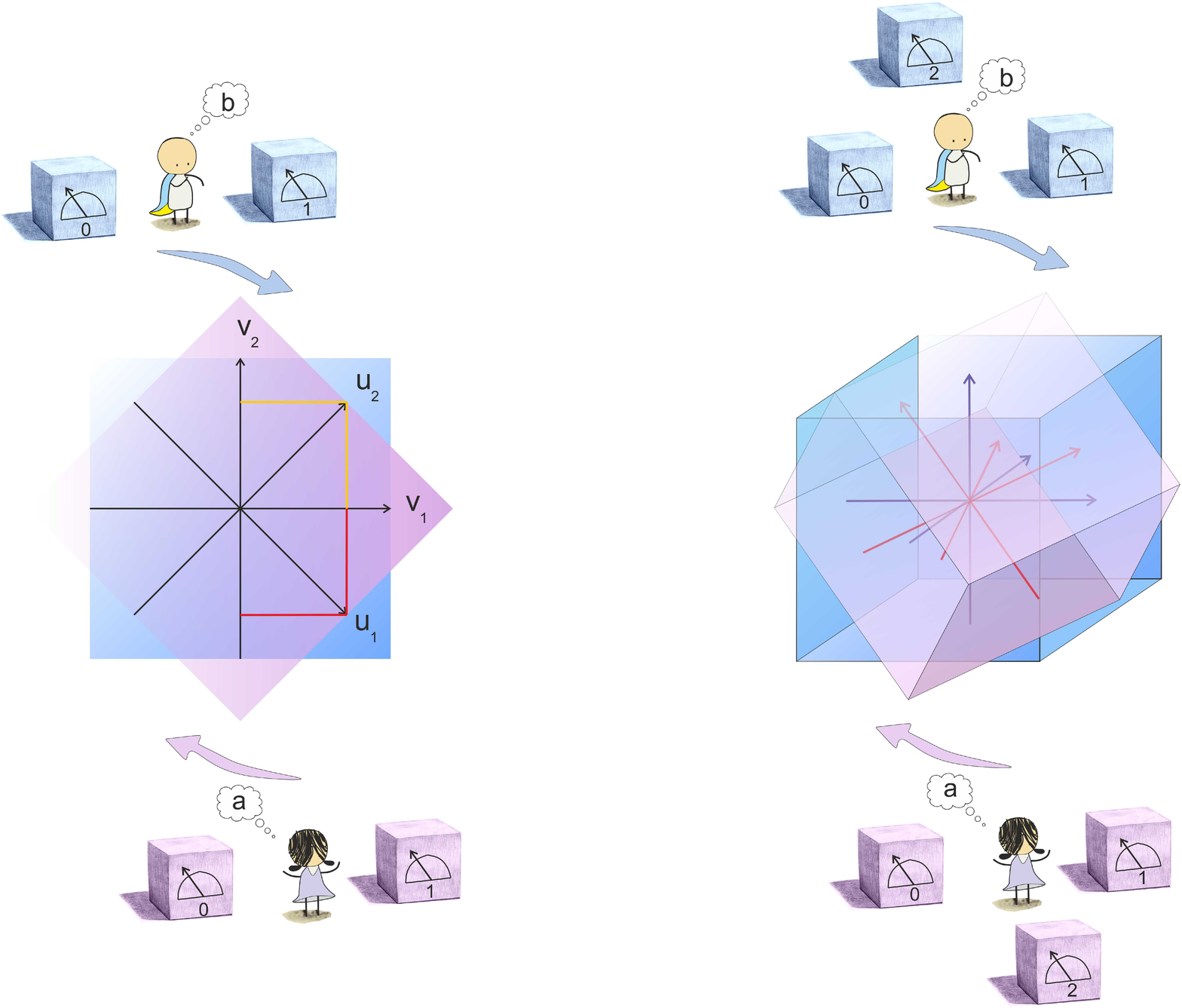}
	
	\caption{The volume maximization game. On the left - a setting in which Alice and Bob each have 2 vectors, and on the right - a setting in which Alice and Bob each have 3 vectors.}
	\label{fig:game}
\end{figure*}

Two remote parties, Alice and Bob, are engaged in a
game in which they attempt to maximize an area of a rectangle (see the left part of Fig. \ref{fig:game}). The
parties are assumed not to communicate by any means.
A single round of the game proceeds as follows. Alice and Bob each have a distinct pair of orthogonal
vectors in $\mathbb{R}^2$,
\begin{equation}\label{key}
\bm{v}_1 = \left[ \begin{array}{c}
1 \\
0
\end{array} \right], \bm{v}_2 = \left[ \begin{array}{c}
0 \\
1
\end{array} \right],
\bm{u}_1 = \left[ \begin{array}{c}
1 \\
-1
\end{array} \right] , \bm{u}_2 = \left[ \begin{array}{c}
1 \\
1
\end{array} \right],
\end{equation}
and a (potentially random) binary variable, $a \in
\{-1,1\}$ for Alice, and $b \in \{-1,1\}$ for Bob. A referee gives Alice one of the vectors she owns, and consequentially she chooses the sign for $a$ and passes both (vector and $a$) to a \emph{walker} that travels in the two-dimensional plane. Bob, on his side does the same:
he gets one of his vectors from the referee, chooses $b$ and passes both to the walker.
Upon receiving the inputs from both players the walker sets its
step size as the length of projection of, say, Alice's vector onto Bob's
vector, and subsequently moves one step forward in
Bob's chosen direction if $a=b$, and one step backward along the same
direction if $a \neq b$. Formally:
\begin{equation}\label{key}
\bm{s}_{ij} \left( a, b \right) = a b \left( \bm{v}_i \cdot \bm{u}_j \right) \bm{u}_j,
\end{equation}
where $ \bm{s}_{ij} \left( a,b \right) $ is the walker's step when Alice uses her $i$th measuring instrument (vector) and Bob uses his $j$th
measuring instrument, for $i,j \in \{1,2\}$.
After a number of rounds the walker's position expressed in Bob's coordinate system (represented by his pair of vectors) defines the following rectangle: one of its vertices is at the origin, the edges that meet at the origin coincide with the axes, and another vertex is at the walker's position. The goal of both parties is to maximize the average area (that is, over many rounds) of this rectangle by choosing the values of $a$, $b$.

Suppose Alice and Bob play a total of $T$ rounds. On each round, $i,j$
are chosen randomly, uniformly and independently of other rounds. Let
$S_{T}$ be a random variable equal to the normalized area after $ T $
rounds,
\begin{equation}\label{key}
S_T = \frac{1}{T^2} \left[ \bm{u}_1 \cdot \sum_{\tau = 1}^T \bm{s} \left( \tau \right) \right] \left[ \bm{u}_2 \cdot \sum_{\tau = 1}^T \bm{s} \left( \tau \right) \right],
\end{equation}
where $ \bm{s} \left( \tau \right) $ is the step vector $ \bm{s}_{ij} \left( a,b \right) $, where $a,b,i,j$ are those generated in the $\tau$th round. Then,
\begin{multline}
E[S_T] = \\
\frac{1}{T^2} E \left[  \sum_{\tau
    : j(\tau)=1} \bm{u}_1 \cdot \bm{s} \left( \tau \right) \right] E
\left[  \sum_{\tau : j(\tau)=2} \bm{u}_2 \cdot \bm{s} \left( \tau
  \right) \right] = \\
\frac{1}{16} \left( c_{11} - c_{21} \right) \left( c_{12} + c_{22}
\right),
\end{multline}
is the average normalized area, where $c_{ij} \triangleq E[ab \vert
i,j]$ is the two-point correlator of Alice and Bob's binary outcomes, $a$ and $b$ respectively,
when Alice uses her $i$th measuring instrument (vector) and Bob uses
his $j$th measuring instrument. Thus, we take our multiplicative
two-device Bell parameter to be:
\begin{equation}
\label{B2}
\mathcal{B}_{2} = \left( c_{11} - c_{21} \right) \left( c_{12}
+ c_{22} \right).
\end{equation}

While the setting of the present game is similar in spirit to the Bell-CHSH
scenario, its objective leads to a whole new class of Bell inequalities whose Bell
parameters are average volumes of hyperrectangles. The above story
describes the simplest of such games with two parties and a
two-dimensional walker.

If in this game Alice and Bob share an EPR state they may cover an average area as twice as large than any classical strategy would allow. In other words, the Bell limit for $ \mathcal{B}_{2} $ is $ 1 $, and the Tsirelson limit is $2$, which can be shown using the well-known inequality of geometric and arithmetic means together with the bounds on the (additive) Bell-CHSH parameter~\cite{Tsirelson, CHSH}. The formal proof can be found in part \textit{I} of the supplementary material~\cite{S-M}.

\section{Multiplicative Bell parameter for $n$ possible measurement choices}
The above game naturally extends to any number of dimensions. The class of Bell inequalities associated with such games is characterized by products of sums of two-point correlators, which are measures of the extent of coordination between Alice and Bob's decisions.
Define
\begin{equation}\label{Bn}
\mathcal{B}_{n} \triangleq  \left(c_{1n}
+\cdots+c_{nn} \right)
\prod_{j=1}^{n-1} \left(c_{1j} +\cdots+ c_{jj}
-j c_{j+1, \, j} \right),
\end{equation}
as the $n$-device \textit{multiplicative Bell parameter}. Geometrically, $\abs{\mathcal{B}_n}$ is proportional to the volume of a particular $n$-dimensional hyperrectangle.
The construction of these parameters is described as follows:
\begin{equation}\label{key}
\mathcal{B}_{n} = \prod_{j=1}^{n} \bm{u}_j \cdot \bm{c}_j ,
\end{equation}
where $ \left\{ \bm{u}_j \mid j = 1,\ldots,n \right\} $ is an orthogonal set of vectors, and $ \bm{c}_j $ is the vector of correlators between Alice's measurement outcomes and Bob's $j$th outcome. A more detailed description of this construction can be found in part \textit{II} of the supplementary material~\cite{S-M}.

As an aside, we note that our multiplicative Bell parameters can be associated with the following additive Bell parameters denoted by $ \mathcal{B}'_n $:
\begin{equation}\label{key}
\mathcal{B}'_n \triangleq \sum_{j=1}^{n} \bm{u}_j \cdot \bm{c}_j
\end{equation}
via the following relation:
\begin{equation}\label{keygg}
\abs{ \mathcal{B}_n } \leq \left( \frac{ \mathcal{B}'_n }{ n }  \right)^n ,
\end{equation}
which is a result of the inequality of geometric and arithmetic means. For $n=2$, this inequality is tight, and $ \mathcal{B}'_2 $ is the well-known Bell-CHSH parameter. More details can be found in part \textit{I} of the supplementary material~\cite{S-M}.

Quantum correlations allow exceeding the classical Bell bound only up
to a certain limit known as the Tsirelson bound~\cite{Tsirelson}. Deriving this quantum
bound is generally a difficult problem~\cite{Bellnonlocality:14}. However, in our case we have the following elegant closed-form expression for the Tsirelson bound, which can be computed efficiently.

\section{Main result}
The Tsirelson bound on the multiplicative $n$-device Bell parameter is $n$ factorial,
\begin{equation}
\abs{\mathcal{B}_{n}} \leq n! \,.
\end{equation}
The bound follows from the positive semi-definiteness of a certain second moment matrix, which, by Schur's complement, yields the following matrix inequality: $ R_A \succeq \bm{c}_j \bm{c}_j^T $. This inequality means that the difference between $R_A$ (Alice's local uncertainty matrix \cite{carmi2018significance}) and the outer product of the correlations vector with itself is a semidefinite positive matrix. A detailed proof, including the players' strategies which saturate this bound, appears in part \textit{III} of the supplementary material~\cite{S-M}.

\section{Bell limit}
The Bell limit is the maximal value of a Bell parameter in local hidden variables theories. Generally, this classical bound cannot be computed efficiently for ordinary (additive) Bell inequalities~\cite{Bellnonlocality:14}. Let us examine \eqref{Bn} while assuming there exists a joint probability distribution for $a_i,b_j$:
\begin{equation}\label{Bn_classical}
\mathcal{B}_{n}=
E\left[b_{n} \sum_{k=1}^n a_k \right]
\prod_{j=1}^{n-1} E\left[b_{j}\left(\sum_{k=1}^j a_k -j a_{j+1}\right)\right],
\end{equation}
where $a_i,b_j$ are random variables corresponding to the values of $a,b$ when Alice and Bob's inputs are $i,j$ respectively.
We suspect that finding a tight bound on \eqref{Bn_classical} is also difficult.
Therefore, we examine the special case where Bob adopts a deterministic strategy. In this case, $\mathcal{B}_n$ is an $n$-variable function of $\mu_i$,
\begin{multline}\label{harmonic}
\mathcal{P}_{n} \left(\mu_1, \ldots, \mu_n
\right) \triangleq \\
\left(
\sum_{i=1}^{n} \mu_i \right) \cdot
\prod_{j=1}^{n-1} \left( \sum_{i=1}^{j} \mu_i - j \cdot \mu_{j+1} \right) ,
\end{multline}
where the local statistics at Alice's site is represented by the
one-point correlators, $\mu_i \triangleq E[a \mid i] ,\, \abs{\mu_i}
\leq 1 $.

As it turns out, $ \mathcal{P}_{n} $ is an $n$-variable harmonic
function. This means that the Laplacian of $ \mathcal{P}_{n} $
vanishes, which implies that its maximum is achieved on the faces of
the n-dimensional hypercube. The proof appears in part \textit{IV} of
the supplementary material~\cite{S-M}. Unfortunately, we have not
succeeded in establishing an efficient way to find the maximum of $
\mathcal{P}_{n} $.

Consider the strategy where Alice and Bob's choices are not only
independent but also deterministic. In this case the values achieved
by $\mathcal{P}_{n}$ constitute a special corner of the $n$-dimensional hypercube:
\begin{equation}\label{FD_strategy}
\mu_{i} = \begin{cases} \begin{array}{c}
\left( -1 \right)^{i} \\
1
\end{array} & \begin{array}{c}
i \le i_c \\
i > i_c
\end{array}
\end{cases}
\end{equation}
i.e., Bob always chooses $ 1 $, and Alice alternates between $ \pm 1 $ until the index reaches some integer cutoff. For indexes larger than this cutoff, Alice always chooses $ 1 $. We shall choose $ i_c $ as the greatest even number which is no larger than $ n - \sqrt{n} $.

\begin{figure}[htb]
	\centering
	
	\includegraphics[width=0.47\textwidth]{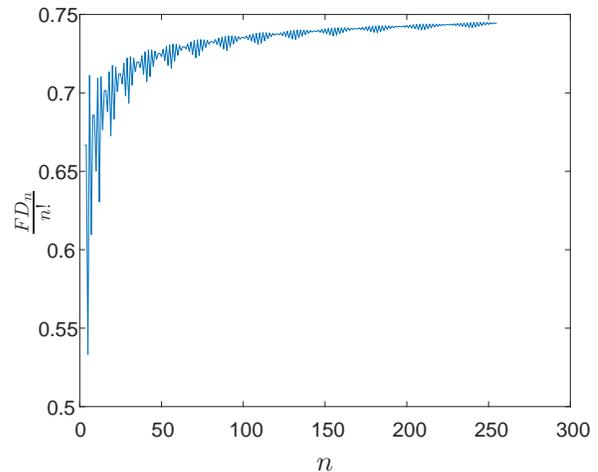}
	
	\caption{\small The ratio between $FD_n$ (lower bound on the
          Bell limit) and $n!$ (Tsirelson's bound) as a function of $n$.}
	\label{fig:ratio}
\end{figure}

We denote the value of $ \mathcal{P}_{n} $ achieved by this strategy as $ FD_{n} $. Since it is a special case of the previous subsection, it is clear that:
\begin{equation}\label{limits}
FD_{n} \le \max P_{n} \left( \vec{\mu} \right) \le \text{Bell limit} \le n!
\end{equation}
Let us write down an analytic expression for $ FD_{n} $. Plugging \eqref{FD_strategy} into \eqref{harmonic}, shows that:
\begin{eqnarray}\label{FD_first}
FD_{n} = \left( 2 \cdot 2 \cdot 4 \cdot 4 \cdots i_c \cdot i_c \right) \cdot \nonumber\\ \underbrace{i_c \cdots i_c}_{n - i_c - 1 \, times}  \cdot \left( n - i_c \right)
\end{eqnarray}
i.e., $ FD_{n} $ is a product of all the even numbers smaller or equal to the cutoff squared, multiplied by the cutoff value ($ n - i_c - 1 $) times, multiplied by the difference between $ n $ and the cutoff value.
It can be shown that \eqref{FD_first} is equivalent to:
\begin{equation}
FD_{n} = 2^{i_c} \cdot \left[ \left( \frac{i_c}{2} \right) ! \right]^{2} \cdot i_c^{n - i_c - 1} \cdot \left( n - i_c \right) .
\end{equation}
As the number of measuring devices grows indefinitely, $ n \rightarrow
\infty $, the ratio between the Bell limit and Tsirelson limit is at
least $ \sqrt{\frac{\pi}{2 e}} $. This result follows from
\eqref{limits} upon noting that
\begin{multline}
\label{eq:pie}
\lim_{n \rightarrow \infty} \frac{FD_{n}}{n !} = \\
\lim_{n \rightarrow \infty} \frac{2^{n - \sqrt{n}} \cdot \left[ \left( \frac{n - \sqrt{n}}{2} \right) ! \right]^{2} \cdot \left( n - \sqrt{n} \right)^{\sqrt{n} - 1} \cdot \sqrt{n} }{n!} = \sqrt{\frac{\pi}{2 e}}
\end{multline}
This ratio is plotted in Fig. \ref{fig:ratio} for values of $n$ up to $255$.

\section{Theoretical and experimental significance}
Linear Bell inequalities are suitable for analyzing local hidden variables, since the latter are described by convex sets created by intersection of linear constraints. However, the quantum set is not a polytope and therefore it can be insightful to study its structure with nonlinear bounds like the one suggested here.

\begin{figure}[htb]
	\centering
	
	\includegraphics[width=0.47\textwidth]{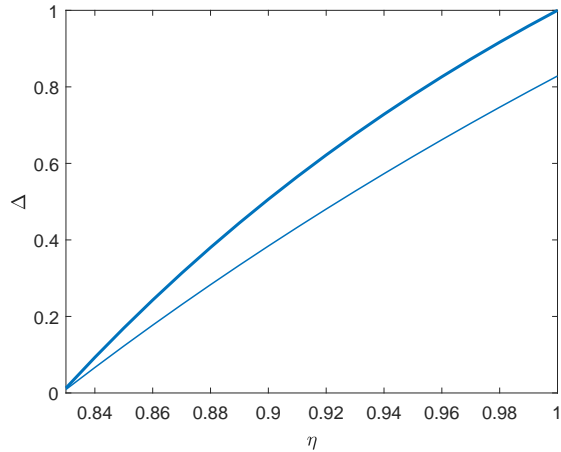}
	
	\caption{\small Robustness of the Bell-CHSH parameter
          $\mathcal{B}_2'$ (thin line) and its multiplicative counterpart
          $\mathcal{B}_2$ (thick line) to the detection loophole.}

	\label{fig:loophole}
\end{figure}

Moreover, the multiplicative bipartite Bell parameter alleviates the
detection loophole in actual Bell experiments. To see how, recall that
in a Bell experiment with photon pairs the detector
efficiency, $\eta$, represents the fraction of incoming photons
registered on the average by the detector. A detector with $\eta = 1$
is perfect but most of the actual detectors have $\eta $ strictly less
than $1$. The detection efficiency influences the classical bound on
the Bell-CHSH parameter, $\abs{\mathcal{B}_2'} \leq 4 / \eta - 2$~\cite{larsson1998bell}. So in actual experiments the Bell bound is always greater
than $2$, and $\mathcal{B}_2'$ becomes virtually ineffective for $\eta = 4 /
(2 \sqrt{2} + 2) \approx 0.83$, when it can no longer discern quantum
behaviors. The robustness of $\mathcal{B}_2'$ to the detection
loophole may be quantified as, $\Delta' = 2 \sqrt{2} -
\left(4 / \eta - 2\right)$, the interval allotted for quantum
violations of the Bell-CHSH inequality. As it turns out, the multiplicative parameter
$\mathcal{B}_2$ exhibits a greater robustness for all values of $\eta$
from $0.83$ to $1$. In particular, according to \eqref{keygg}, $\abs{\mathcal{B}_2} \leq (2 / \eta - 1)^2$, and therefore
its robustness is $\Delta = 2 - (2 / \eta - 1)^2$, which as shown
in Figure~\ref{fig:loophole} is always greater than $\Delta'$ for all values of $\eta \in
(0.83, 1]$.

\section{Simulation}
\begin{figure}[htb]
	\centering
	
	\includegraphics[width=0.49\textwidth]{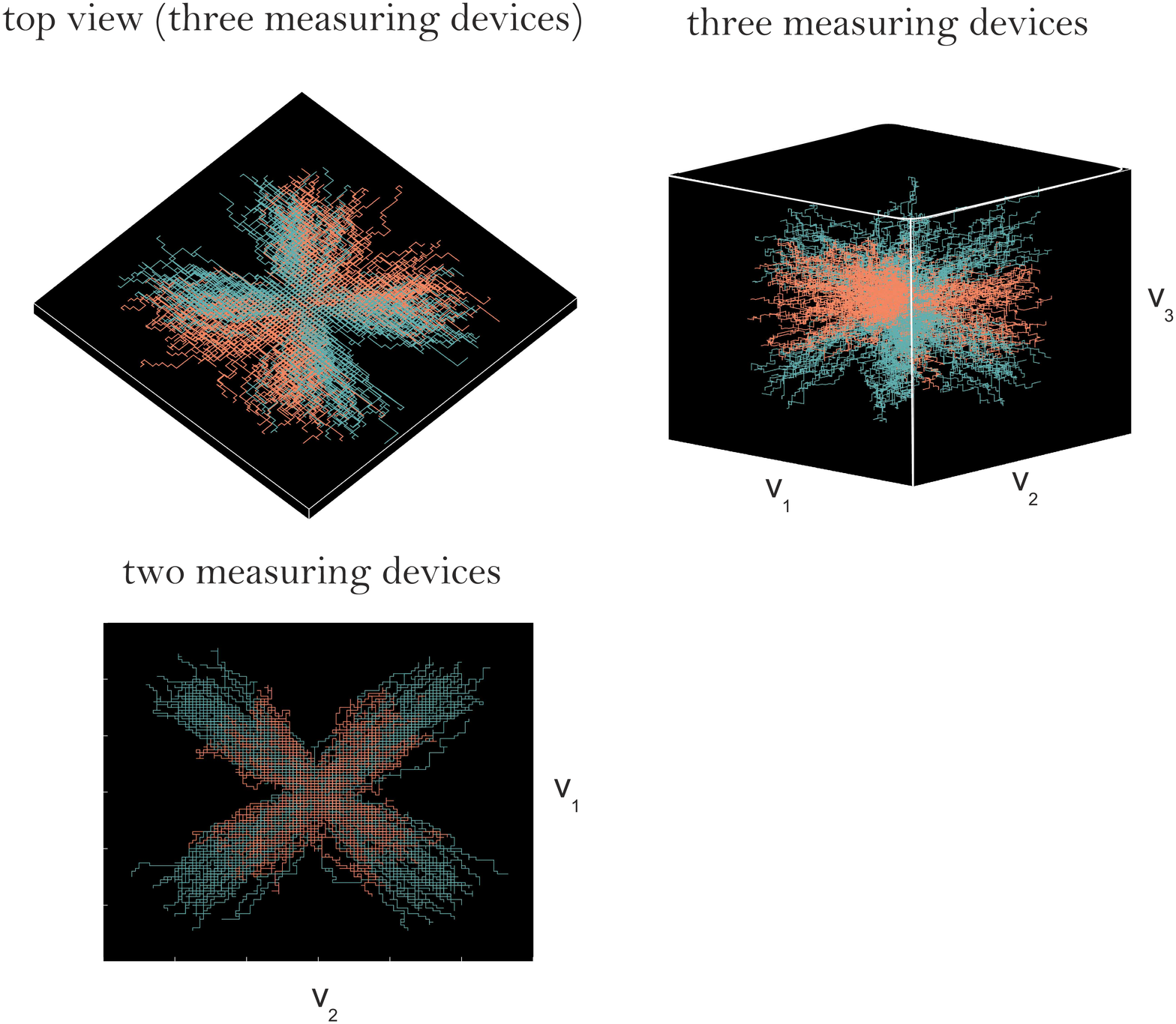}
	
	\caption{\small The walker's random paths in two and three
		dimensions (corresponding to the number of measuring
                instruments). The orange paths correspond to classical
                strategies, while the cyan paths correspond to the
                ``winning'' quantum strategies. Clearly, the
                statistics of the quantum paths differ from the
                classical ones. This premise alludes to a test of
                quantumness where the covariance or some other
                statistic is empirically evaluated over the
                paths generated in different trials of the volume
                maximization game.}

	\label{fig:walker}
\end{figure}

We ran simulations of the volume maximization game for $n=2,3$ (see Fig. \ref{fig:walker}). For the classical strategy in $n=2$, we used: $ a_1 = b_1 = b_2 = 1, E \left[ a_2 \right]=0 $ (optimality of this strategy is proven in the supplementary material - see part \textit{I}). For the classical strategy $n=3$, we used the fully-deterministic strategy described in the former paragraph. For the quantum strategies in both $n=2,3$, we used the ``winning'' strategies described in the proof of our main result.

\section{Conclusions}
We have explored a new coordination game which favors quantum players. This has allowed us to find a new class of multiplicative Bell inequalities, as well as their corresponding Tsirelson bounds.
It was shown that in quantum mechanics any $n$-device multiplicative Bell
parameter is bounded by the volume of an $n$-dimensional ellipsoid
representing the local uncertainty associated with the system of one
of the players. Contrary to our intuition, as $n$ goes to infinity, the performances of classical and quantum players become comparable.

Some future generalizations of our game and associated parameter may include: multiple players; allowing non-binary measurement outcomes, both discrete and continuous; using a different set of vectors $\bm{u}_j$ for the construction, possibly a non-orthogonal one. Another research direction might include a deeper investigation of the associated additive Bell parameters $\mathcal{B}'_n$, and their relation to the multiplicative ones.
Finally, stronger-than quantum correlations within post-quantum models
may be analyzed using the current approach.

\section*{Acknowledgments}
We thank Pawel Horodecki for insightful discussions. A.C. acknowledges support from the Israel Science Foundation, Grant
No. 1723/16. E.C. was supported by the Canada Research Chairs (CRC)
Program and by the Engineering Faculty in Bar Ilan University.


\bibliographystyle{apsrev4-1}
\bibliography{ArticleReview}

\begin{widetext}
\newpage
\appendix

\section*{Supplementary Material}
\renewcommand{\theparagraph}{\Roman{paragraph}}
\paragraph{Bell and Tsirelson bounds for the case of $n=2$}

\begin{proof}
	From the inequality of geometric and arithmetic means:
	\begin{equation}\label{key}
	\mathcal{B}_{2} = \left( c_{11} - c_{21} \right) \left( c_{12} + c_{22} \right) \leq \left( \frac{c_{11} - c_{21} + c_{12} + c_{22} }{2} \right)^2
	\end{equation}
	where on the LHS we have $\mathcal{B}_2$, and on the RHS the numerator is equivalent to the well-known Bell-CHSH parameter. By plugging the known Bell and Tsirelson bounds for the Bell-CHSH parameter~\cite{CHSH,Tsirelson}, it can be shown that the respective bounds for $\mathcal{B}_2$ are $1$ and $2$. In order to show that these bounds can be saturated, let us demonstrate a local hidden-variable strategy:
	$ a_1 = b_1 = b_2 = 1, E \left[ a_2 \right]=0 $. And the quantum strategy should be same as the one which saturates Tsirelson's bound for Bell-CHSH, where $c_{ij}=\frac{1}{\sqrt{2}} \left( 1 - 2 \delta_{i,2} \delta_{j,1} \right) $.
\end{proof}

In general, our $n$-device multiplicative Bell parameter can be associated with an additive one in a similar way. Again, from the inequality of geometric and arithmetic means:
\begin{equation}\label{AM-GM}
\sqrt[n]{\mathcal{B}_n} = \sqrt[n]{ \prod_{j=1}^{n} \bm{u}_j \cdot \bm{c}_j } \leq \frac{\sum_{j=1}^{n} \bm{u}_j \cdot \bm{c}_j}{n}
\end{equation}
Thus, let us define the associated $n$-device additive Bell parameter as:
\begin{equation}\label{Assoc}
\mathcal{B}'_n \triangleq \sum_{j=1}^{n} \bm{u}_j \cdot \bm{c}_j
\end{equation}
and by plugging \eqref{Assoc} into \eqref{AM-GM}, we have:
\begin{equation}\label{key}
\abs{ \mathcal{B}_n } \leq \left( \frac{ \mathcal{B}'_n }{n } \right)^n
\end{equation}

\paragraph{Construction of the Multiplicative Bell Parameter for n measurement devices}

Let $ X $ be the following vector of quantum operators:
\begin{equation}\label{key}
X=\left[\begin{array}{c}
B_{j}\\
A_{1}\\
\vdots\\
A_{n}
\end{array}\right]
\end{equation}
where $ i,j\in\left\{ 1,2,\ldots,n\right\} $ are Alice and Bob's inputs, and $ A_{i},B_{j} $  are operators with spectrum $\left\{ \pm 1 \right\} $, which represent the outcomes of Alice and Bob's measurements, respectively.


The second moment matrix of $ X $ is defined by the following relation:
\begin{equation}\label{cov_n}
\varSigma_{ij} = \braket{ X_{i}X_{j} }.
\end{equation}


It immediately follows that
$ \braket{ B_{j}B_{j} }=\braket{ A_{i}A_{i} }=1 $.
Since $ A_{i},A_{j} $ are non-commuting linear operators (for $ i \neq j $), the expected value $ \braket{A_{i}A_{j}} $ cannot be measured in any experiment, and is generally a complex number. We shall denote it by $ r_{ij} $:
\begin{equation}\label{r_ij}
r_{ij}\triangleq\left\langle A_{i}A_{j}\right\rangle =\left\langle A_{j}A_{i}\right\rangle ^{*}.
\end{equation}

The expected values of $ A_{i}B_{j} $ are the aforementioned two-point correlators between Alice and Bob's measurement results, and had been denoted by $ c_{ij} $:
\begin{equation}\label{cor_ij}
c_{ij} \triangleq \braket{A_i \otimes B_j}
\end{equation}

note that $c_{ij} \in\left[-1,1\right] $.

Therefore, the second moment matrix $ \varSigma $ is equal to:
\begin{eqnarray}\label{key}
\varSigma=
\left[\begin{array}{ccccc}
1 & c_{1j} & c_{2j} & \cdots & c_{nj}\\
c_{1j} & 1 & r_{12} & \cdots & r_{1n}\\
c_{2j} & r_{21} & 1 & \ddots & \vdots\\
\vdots & \vdots & \ddots & \ddots & r_{n-1,n}\\
c_{nj} & r_{n1} & \cdots & r_{n,n-1} & 1
\end{array}\right]\succeq0
\end{eqnarray}
and it is positive semi-definite~\cite{carmi2018significance}.

Using Schur's complement, we obtain:
\begin{equation}\label{Schur_n}
\mathcal{R}_{A}\triangleq\left[\begin{array}{ccc}
1 & \cdots & r_{ij}\\
\vdots & \ddots & \vdots\\
r_{ji} & \cdots & 1
\end{array}\right]\succeq\left[\begin{array}{c}
c_{1j}\\
\vdots\\
c_{nj}
\end{array}\right]\left[\begin{array}{ccc}
c_{1j} & \cdots & c_{nj}\end{array}\right]
\end{equation}

We denote:
\begin{equation}\label{key}
\bm{c}_{j} \triangleq \left[\begin{array}{c}
c_{1j}\\
\vdots\\
c_{nj}
\end{array}\right], \mathcal{C}_{j} \triangleq \bm{c}_{j} \bm{c}_{j}^{T}
\end{equation}
Now, let us construct the following $ n \times n $ matrix:
\begin{equation}\label{R_matrix}
\varGamma \triangleq\left[\begin{array}{cccc}
1 & r & \cdots & r\\
r & 1 & r & \vdots\\
\vdots & r & \ddots & r\\
r & \cdots & r & 1
\end{array}\right]
\end{equation}
where $ r \in \mathbb{R} $. Note that $ \varGamma $ is real and symmetric. Its orthogonal eigenvectors are:
\begin{eqnarray*}\label{key}
	\bm{u}_{j}\left(l\right)=\begin{cases}
		\begin{array}{c}
			1\\
			-j\\
			0
		\end{array} & \begin{array}{c}
			l\leq j\\
			l=j+1\\
			l>j+1
	\end{array}\end{cases}\,\,for\,1\leq j<n,\,\bm{u}_{n}=\left[\begin{array}{c}
		1\\
		\vdots\\
		1
	\end{array}\right]
\end{eqnarray*}
Now, we shall use the vectors $ \bm{u}_{k} $ in order to construct our multiplicative Bell parameter:
\begin{equation}\label{Bell_multiplicative_n}
\mathcal{B}_{n} \triangleq \prod_{j=1}^{n} \bm{c}_{j} \cdot \bm{u}_{j}
\end{equation}

\paragraph{Proof of main result}
In order to prove this result, it is required to prove two parts:
\begin{enumerate}[label=(\alph*)]
	\item\label{fact_inequality} $ \abs{ \mathcal{B}_n } \leq n! $
	\item\label{fact_equality} $ \abs{ \mathcal{B}_n } = n! $ can be achieved in quantum mechanics.
\end{enumerate}
\begin{proof}[Proof \ref{fact_inequality}]
	First, let us denote: $ \hat{u}_{j} \triangleq \frac{\bm{u}_{j}}{\left\| \bm{u}_{j} \right\|} $.
	$ \mathcal{B}_n \leq n! $ follows from:
	\begin{enumerate}
		\item\label{proof_factorial} $ \prod_{j=1}^{n} \left\| \bm{u}_{j} \right\| = n! $
		\item\label{proof_schur} $ \forall j \in \{1,2,\ldots n\}, \bm{u}_{j}^{T} c_{j} c_{j}^{T} \bm{u}_{j} \leq \bm{u}_{j}^{T} \mathcal{R}_{A} \bm{u}_{j} $
		\item\label{proof_mean} $ \prod_{j=1}^{n} \hat{u}_{j}^{T} \mathcal{R}_{A} \hat{u}_{j} \leq \left( \frac{\sum_{j=1}^{n} \hat{u}_{j}^{T} \mathcal{R}_{A} \hat{u}_{j} }{n} \right)^{n} $
		\item\label{proof_hard_part} $ \sum_{j=1}^{n} \hat{u}_{j}^{T} \mathcal{R}_{A} \hat{u}_{j} = n $
	\end{enumerate}
	Let us calculate the product of all the eigenvectors' squared norms:
	\begin{equation}\label{norm_prod}
	\left\Vert \bm{u}_{n}\right\Vert ^{2}\prod_{k=1}^{n-1}\left\Vert \bm{u}_{k}\right\Vert ^{2}=n\prod_{k=1}^{n-1}\left(k+k^{2}\right)=\left(n!\right)^{2}
	\end{equation}
	which proves \ref{proof_factorial}. \ref{proof_schur} is a direct result of \eqref{Schur_n}, and \ref{proof_mean} follows immediately from the inequality of arithmetic and geometric means. So all we have left is to prove \ref{proof_hard_part}:
		\begin{eqnarray*}\label{key}
			&& \sum_{j=1}^{n} \hat{u}_{j}^T \mathcal{R}_A \hat{u}_j = \sum_{j=1}^{n} \sum_{l=1}^{n} \sum_{k=1}^{n} \hat{u}_{j} \left( l \right) \mathcal{R}_A \left( l, k \right) \hat{u}_j \left( k \right) = \sum_{j=1}^{n-1} \sum_{l=1}^{n} \sum_{k=1}^{n} \hat{u}_{j} \left( l \right) r_{lk} \hat{u}_j \left( k \right) + \sum_{l=1}^{n} \sum_{k=1}^{n} \hat{u}_{n} \left( l \right) r_{lk} \hat{u}_n \left( k \right) = \\
			&& \sum_{j=1}^{n-1} \frac{1}{ \left\| \bm{u}_j \right\|^2 } \left[ \sum_{l=1}^{j} \left( \sum_{k=1}^{j} u_{j} \left( l \right) r_{lk} u_j \left( k \right) + u_{j} \left( l \right) r_{l,j+1} u_j \left( j+1 \right) \right) + \sum_{k=1}^{j} u_{j} \left( j+1 \right) r_{j+1,k} u_j \left( k \right) + \left[ u_{j} \left( j+1 \right) \right]^2 \right] + \frac{1}{n} \sum_{l=1}^{n} \sum_{k=1}^{n} r_{lk} = \\
			&& \sum_{j=1}^{n-1} \frac{1}{ j \left( j + 1 \right) } \left[ \sum_{l=1}^{j} \left( \sum_{k=1}^{j} r_{lk} - j r_{l,j+1} \right) - \sum_{k=1}^{j} j r_{j+1,k} + j^2 \right] + \frac{1}{n} \left[ \sum_{l=1}^{n} 1 + \sum_{ 1 \leq l < k \leq n} \left( r_{lk} + r_{kl} \right) \right] = \\
			&& \sum_{j=1}^{n-1} \frac{1}{ j \left( j + 1 \right) } \left[ j + 2 \sum_{ 1 \leq l < k \leq j} Re \left\{ r_{lk} \right\} - j \sum_{l=1}^{j} r_{l, j+1} - j \sum_{k=1}^{j} r_{j+1,k} + j^2 \right] + \frac{1}{n} \left[ n + 2 \sum_{ 1 \leq l < k \leq n} Re \left\{ r_{lk} \right\} \right] = \\
			&& n - 1 + \sum_{j=1}^{n-1} \frac{1}{ j \left( j + 1 \right) } \left[ 2 \sum_{ 1 \leq l < k \leq j} Re \left\{ r_{lk} \right\} - 2 j \sum_{l=1}^{j} Re \left\{ r_{l, j+1} \right\} \right] + 1 + \frac{2}{n} \sum_{ 1 \leq l < k \leq n} Re \left\{ r_{lk} \right\} = \\
			&& n + 	\sum_{j=1}^{n-1} \frac{2}{ j \left( j + 1 \right) } \left[ \sum_{ 1 \leq l < k \leq j} Re \left\{ r_{lk} \right\} - j \sum_{l=1}^{j} Re \left\{ r_{l, j+1} \right\} \right] + \frac{2}{n} \sum_{ 1 \leq l < k \leq n} Re \left\{ r_{lk} \right\} = \\
			&& n + \sum_{ 1 \leq l < k \leq n-1 } Re \left\{ r_{lk} \right\} \left[ \sum_{j=k}^{n-1} \frac{2}{ j \left( j + 1 \right) } - \frac{2}{k} + \frac{2}{n} \right] + \sum_{ l=1 }^n Re \left\{ r_{ln} \right\} \left( - \frac{2}{n} + \frac{2}{n} \right) = n
		\end{eqnarray*}
where the transitions follow by computing the coefficient of $ Re \left\{ r_{lk} \right\} $ for each pair $l,k$ s.t. $ l < k $, where we separated the cases of $ k < n $ and $ k = n $. The final transition uses the sum: $ \sum_{j=1}^{n} \frac{1}{j \left( j+1 \right)} = \frac{n}{n+1} $.
	
	To summarize: \ref{proof_mean} and \ref{proof_hard_part} show that $ \prod_{j=1}^{n} \hat{u}_j^T \mathcal{R}_A \hat{u}_j \leq 1 $, which when combined with \ref{proof_schur} implies:
	\begin{equation}\label{proof_almost_done}
	1 \geq \prod_{j=1}^{n} \left| c_{j} \cdot \hat{u}_{j} \right|^{2} = \prod_{j=1}^{n} \frac{1}{ \left\| \bm{u}_{j} \right\|^{2} } \prod_{j=1}^{n} \left| c_{j} \cdot \bm{u}_{j} \right|^{2}
	\end{equation}
	Finally, substituting \ref{proof_factorial} into \eqref{proof_almost_done} gives us \ref{fact_inequality}.
\end{proof}
\begin{proof}[Proof \ref{fact_equality}]
	First, let us assume that Alice and Bob each have a qubit, and their outputs $ A_i, B_j $ are results of measurements on their respective qubits. Following this assumption, we begin by computing the quantum expected value of $ A_{i} A_{k} $, where $ \hat{a}_{i} $ is a normalized vector which signifies Alice's $ i $th measurement direction, i.e., $ A_{i} = \hat{a}_{i}\cdot \vec{\sigma} $. We denote Alice's density matrix as $ \rho_{A}$.
	\begin{eqnarray*}
		\left\langle A_{i} A_{k} \right\rangle =  tr \left( \rho_{A} A_{i} A_{k} \right) = tr \left[ \rho_{A} \left(\hat{a}_{i}\cdot \vec{\sigma}\right) \left(\hat{a}_{k}\cdot \vec{\sigma}\right) \right] =
		tr \left[ \left(\hat{a}_{i}\cdot\hat{a}_{k}\right) \rho_{A}+i \rho_{A} \left( \hat{a}_{i} \times \hat{a}_{k} \right) \cdot \vec{\sigma} \right] =
		\hat{a}_{i}\cdot\hat{a}_{k} + i \cdot tr \left[ \rho_{A} \left( \hat{a}_{i} \times \hat{a}_{k} \right) \cdot \vec{\sigma} \right]
	\end{eqnarray*}
	Note that $ tr \left[ \rho_{A} \left( \hat{a}_{i} \times \hat{a}_{k} \right) \cdot \vec{\sigma} \right] $ is real, meaning that it is the imaginary part of $ \left\langle A_{i} A_{k} \right\rangle $. Thus, for any state, the matrix $ \mathcal{R}_{A} $ is as follows:
	\begin{equation}\label{R_A_ik}
	\left( \mathcal{R}_{A} \right)_{ik} = \hat{a}_{i} \cdot \hat{a}_{k} + i \cdot tr \left[ \rho_{A} \left( \hat{a}_{i} \times \hat{a}_{k} \right) \cdot \vec{\sigma} \right]
	\end{equation}
	Let us denote:
	\begin{equation}\label{key}
	\mathcal{V}_{A}\triangleq\left[\begin{array}{ccc}
	\mid &  & \mid\\
	\hat{a}_{1} & \cdots & \hat{a}_{n}\\
	\mid &  & \mid
	\end{array}\right]
	\end{equation}
	and also $ T_{A} $ shall be a $ n \times n $ matrix, s.t.
	\begin{equation*}\label{key}
	\left( T_{A} \right)_{ik} \triangleq Im \left\{ \left( \mathcal{R}_{A} \right)_{ik} \right\} = tr \left[ \rho_{A} \left( \hat{a}_{i} \times \hat{a}_{k} \right) \cdot \vec{\sigma} \right] .
	\end{equation*}
	Note that $ T_A $ is anti-symmetric (this follows immediately from anti-commutativity of the vector product). From the last two definitions, we have:
	\begin{equation}\label{R_A_decomposition}
	\mathcal{R}_{A} = \mathcal{V}_{A}^{T} \mathcal{V}_{A} + i T_{A}
	\end{equation}
	
	Our proof for \ref{fact_inequality}, also shows that the quantum limit is reached if and only if the following two inequalities are saturated:
	\begin{enumerate}
		\item\label{Alice_inequality} $ \prod_{j=1}^{n} \hat{u}_j^T \mathcal{R}_A \hat{u}_j \leq 1 $
		\item\label{Bob_inequality} $ \forall j \in \{ 1, 2, \ldots n \} , \left( \bm{c}_{j} \cdot \bm{u}_{j} \right)^2 \leq \bm{u}_j^T \mathcal{R}_A \bm{u}_j $
	\end{enumerate}
	Here we will show that Alice can always choose her measurement directions $ \{ \hat{a}_{i}  \} $ s.t. \ref{Alice_inequality} is saturated, and Bob can always choose his measurement directions $ \{ \hat{b}_{j} \} $ s.t. \ref{Bob_inequality} is saturated.
	
	We start by proving the following identity for every vector $ \bm{u} \in \mathbb{R}^n $:
	\begin{equation}\label{T_A_irrelevant}
	\bm{u}^T \mathcal{R}_A \bm{u} = \bm{u}^{T} \mathcal{V}_{A}^{T} \mathcal{V}_{A} \bm{u} + i \bm{u}^{T} T_{A} \bm{u} = \left\| \mathcal{V}_{A} \bm{u} \right\|^{2}
	\end{equation}
	the last transition follows from $ \bm{u}^{T} T_{A} \bm{u} = 0 $, which is true for all real vectors $ \bm{u} $ since $ T_{A} $ is anti-symmetric.
	
	Let us show that:
	\begin{equation}\label{Alice_construct}
	\exists \mathcal{V}_A : \left\| \mathcal{V}_{A} \bm{u}_{j} \right\|^{2} = \left\| \bm{u}_{j} \right\|^{2} ,\, \forall j \in \left[ n \right]
	\end{equation}
	For $ j \neq n $:
	\begin{eqnarray*}\label{key}
		&&	\left\| \mathcal{V}_{A} \bm{u}_{j} \right\|^{2} = \left( \mathcal{V}_{A} \bm{u}_{j} \right) \cdot \left( \mathcal{V}_{A} \bm{u}_{j} \right) =
		\left( \sum_{i=1}^{n} \bm{u}_{j} \left(i\right) \hat{a}_{i} \right) \cdot \left( \sum_{k=1}^{n} \bm{u}_{j} \left(k\right) \hat{a}_{k} \right) =
		\sum_{i=1}^{n} \sum_{k=1}^{n} \bm{u}_{j} \left(i\right) \bm{u}_{j} \left(k\right) \hat{a}_{i} \cdot \hat{a}_{k} = \\
		&& \left\| \bm{u}_{j} \right\|^{2} + 2 \sum_{i=1}^{j} \left[ \left( \sum_{k=i+1}^{j} \hat{a}_{i} \cdot \hat{a}_{k} \right) - j \hat{a}_{i} \cdot \hat{a}_{j+1} \right] =
		\left\| \bm{u}_{j} \right\|^{2} + 2 \sum_{i=1}^{j} \hat{a}_{i} \cdot \left( \sum_{k=i+1}^{j} \hat{a}_{k} - j \hat{a}_{j+1} \right)
	\end{eqnarray*}
	and for $ j = n $:
	\begin{equation}\label{key}
	\left\| \mathcal{V}_{A} \bm{u}_{n} \right\|^{2} = \left\| \bm{u}_{n} \right\|^{2} + 2 \sum_{i=1}^{n-1} \sum_{k=i+1}^{n} \hat{a}_{i} \cdot \hat{a}_{k}
	\end{equation}
	which implies that in order to obtain the equalities $ \left\| \mathcal{V}_{A} \bm{u}_{j} \right\|^{2} = \left\| \bm{u}_{j} \right\|^{2} $ for all $ j \in \left[ n \right] $, it is sufficient to choose $ \{ a_{i} \}$ s.t.:
	\begin{eqnarray}\label{Alice_conditions}
	\begin{cases}
	\forall j \in \left[ n-1 \right], \sum_{i=1}^{j} \hat{a}_{i} \cdot \left( \sum_{k=i+1}^{j} \hat{a}_{k} - j \hat{a}_{j+1} \right) = 0 \\
	\sum_{i=1}^{n-1} \sum_{k=i+1}^{n} \hat{a}_{i} \cdot \hat{a}_{k} = 0
	\end{cases}
	\end{eqnarray}
	this can be achieved as follows:
	\begin{enumerate}[label=(\roman*)]
		\item Choose $ \hat{a}_{1} $ arbitrarily.
		\item For each $ i = 2,3,\ldots, n $, choose $ \hat{a}_{i} $ which is orthogonal to the sum: $ \sum_{j=1}^{i-1} \hat{a}_{j} $
	\end{enumerate}
	Let us prove that this construction satisfies \eqref{Alice_conditions} by induction with respect to $ j $.
	
	\textbf{Basis - $j=1$:} $ \hat{a}_{1} \cdot \left( - \hat{a}_{2} \right) = 0 $
	
	\textbf{Step} - we assume the claim is satisfied for $ j $ and prove it for $ j+1 $:
\begin{eqnarray*}\label{key}
	&&\sum_{i=1}^{j+1} \hat{a}_{i} \cdot \left( \sum_{k=i+1}^{j+1} \hat{a}_{k} - \left(j+1\right) \hat{a}_{j+2} \right) =
			\sum_{i=1}^{j+1} \hat{a}_{i} \cdot \left( \sum_{k=i+1}^{j+1} \hat{a}_{k} \right) - \left(j+1\right) \left( \sum_{i=1}^{j+1} \hat{a}_{i} \right) \cdot \hat{a}_{j+2} =
			\sum_{i=1}^{j+1} \hat{a}_{i} \cdot \left( \sum_{k=i+1}^{j+1} \hat{a}_{k} \right) =\\&& \sum_{i=1}^{j} \hat{a}_{i} \cdot \left( \sum_{k=i+1}^{j+1} \hat{a}_{k} \right) = \sum_{i=1}^{j} \sum_{k=i+1}^{j+1} \hat{a}_{i} \cdot \hat{a}_{k} =
			\sum_{i=1}^{j} \hat{a}_{i} \cdot \left( \sum_{k=i+1}^{j} \hat{a}_{k} \right) + \left( \sum_{i=1}^{j} \hat{a}_{i} \right) \cdot \hat{a}_{j+1} = \sum_{i=1}^{j} \hat{a}_{i} \cdot \left( \sum_{k=i+1}^{j} \hat{a}_{k} \right) = \\&&
			\sum_{i=1}^{j} \sum_{k=i+1}^{j} \hat{a}_{i} \cdot \hat{a}_{k} - j \sum_{i=1}^{j} \hat{a}_{i} \cdot \hat{a}_{j+1} = \sum_{i=1}^{j} \hat{a}_{i} \cdot \left( \sum_{k=i+1}^{j} \hat{a}_{k} - j \hat{a}_{j+1} \right) = 0
		\end{eqnarray*}
note that this proof also proves the condition for $ n $, by substituting $ j = n-1 $ in the simplest expression $ \sum_{i=1}^{j} \sum_{k=i+1}^{j+1} \hat{a}_{i} \cdot \hat{a}_{k} $.
	Combining \eqref{T_A_irrelevant} with \eqref{Alice_construct} implies that indeed, Alice can choose her measurements in a way that always saturates \ref{Alice_inequality}.
	
	In order to show how Bob can saturate \ref{Bob_inequality}, we shall thoroughly examine our system, assuming Alice and Bob share the following quantum state $ \left|\psi\right\rangle $:
	\begin{equation}\label{key}
	\left|\psi\right\rangle =\frac{1}{\sqrt{2}}(\left|00\right\rangle +\left|11\right\rangle )
	\end{equation}
	which is actually the known $\left|\beta_{0}\right\rangle  $ Bell state.
	
	Let us compute the correlations $ c_{ij} $:
	\begin{eqnarray}\label{key}
	c_{ij} && = \left\langle \psi \left| A_{i} \otimes B_{j} \right| \psi \right\rangle = \nonumber\\
	&&	= \frac{1}{2} \left[ \left\langle 0 \left| A_{i} \right| 0 \right\rangle \left\langle 0 \left| B_{j} \right| 0 \right\rangle +  \left\langle 1 \left| A_{i} \right| 0 \right\rangle \left\langle 1 \left| B_{j} \right| 0 \right\rangle
	+ \left\langle 0 \left| A_{i} \right| 1 \right\rangle \left\langle 0 \left| B_{j} \right| 1 \right\rangle +  \left\langle 1 \left| A_{i} \right| 1 \right\rangle \left\langle 1 \left| B_{j} \right| 1 \right\rangle \right]
	= \frac{1}{2} tr \left( A_{i}^{T} B_{j} \right)
	\end{eqnarray}

	Where $ B_{j} = \hat{b}_{j} \cdot \vec{\sigma} $. Note that the expression $ \frac{1}{2} tr \left( A_{i}^{T} B_{j} \right) $ is similar to the Frobenius inner product; however, for complex matrices the Hermitian conjugate of the first matrix should be taken (rather than its transpose):
	\begin{equation*}\label{key}
	\left\langle A,B \right\rangle_{F} \triangleq tr \left( A^{\dagger} B \right) = tr \left( \overline{{A}^{T}} B \right)
	\end{equation*}
	where the overline denotes (element-wise) complex conjugation. By replacing order of transposition and conjugation, we have:
	\begin{equation*}\label{key}
	\left\langle \overline{A},B \right\rangle_{F} = tr \left( \overline{A^{\dagger}} B \right) = tr \left( {A}^{T} B \right)
	\end{equation*}
	which implies that $ c_{ij} = \frac{1}{2} \left\langle \overline{A_{i}},B_{j} \right\rangle_{F} $. We take a closer look at $ \overline{A_{i}} $:
	\begin{equation*}\label{key}
	\overline{A_{i}} = \overline{\hat{a}_{i} \cdot \vec{\sigma}} = \hat{a}_{i} \cdot \overline{\vec{\sigma}}
	\end{equation*}
	since $ \hat{a}_{i} $ is a real vector. The Pauli matrices $ \sigma_{x}, \sigma_{z} $ are also real, and $ \sigma_{y} $ is purely imaginary, so $ \overline{\sigma_{y}} = - \sigma_{y} $, and it follows that $ \overline{A_{i}} = R_{y} \hat{a}_{i} \cdot \vec{\sigma} $, where:
	\begin{equation}\label{R_y}
	R_{y} \triangleq \left[\begin{array}{ccc}
	1 & 0 & 0 \\
	0 & -1 & 0 \\
	0  & 0 & 1
	\end{array}\right]
	\end{equation}
	$ R_{y} $ is a real, orthogonal symmetric which flips the sign of a vector's y component, i.e., $ R_{y} $ is a reflection relative to the $ X-Z $ plane. Using $ R_{y} $, we shall obtain our final expression for $ c_{ij} $:
	\begin{eqnarray}\label{key}
	c_{ij} = \frac{1}{2} tr \left[ \left( R_{y} \hat{a}_{i} \cdot \vec{\sigma} \right) \left( \hat{b}_{j} \cdot \vec{\sigma} \right) \right] =
	\left( R_{y} \hat{a}_{i} \right) \cdot \hat{b}_{j}  = \hat{a}_{i} \cdot \left( R_{y} \hat{b}_{j} \right)
	\end{eqnarray}
	i.e., for $ | \psi \rangle = | \beta_{0} \rangle $, the correlation $ c_{ij} $ is simply the dot product between Alice and Bob's respective measurement directions, with one of them reflected relative to the $ X - Z $ plane. We use this in order to find an expression for the $ \bm{c}_{j} $ vector.
	\begin{eqnarray}\label{c_j_beta_0}
	\bm{c}_{j} \triangleq \left[\begin{array}{c}
	c_{1j}\\
	\vdots\\
	c_{nj}
	\end{array}\right] =
	\left[\begin{array}{c}
	\hat{a}_{1} \cdot R_{y} \hat{b}_{j} \\
	\vdots\\
	\hat{a}_{n} \cdot R_{y} \hat{b}_{j}
	\end{array}\right] =
	\left[\begin{array}{ccc}
	\hbox{---} & \hat{a}_{1} & - \\
	& \vdots & \\
	\hbox{---} &  \hat{a}_{n} & -
	\end{array}\right]
	R_{y}
	\left[\begin{array}{c}
	\mid \\
	\hat{b}_{j} \\
	\mid
	\end{array}\right] = \mathcal{V}_{A}^{T} R_{y} \hat{b}_{j}
	\end{eqnarray}
	By recalling the way we constructed our Bell parameter, it follows that:
	\begin{eqnarray}\label{key}
	\mathcal{B}_{n}  = \prod_{j=1}^{n} \bm{u}_j \cdot \bm{c}_j
	= \prod_{j=1}^{n} \bm{u}_{j} \cdot \mathcal{V}_{A}^{T} R_{y} \hat{b}_{j}  =
	\prod_{j=1}^{n} \hat{b}_{j} \cdot \left( R_{y} \mathcal{V}_{A} \bm{u}_{j} \right)
	\end{eqnarray}
	which implies that for a given $ \mathcal{V}_{A} $, the following choice for Bob's measurement directions would maximize $ \mathcal{B}_{n} $:
	\begin{equation}\label{Bob_optimal_choices}
	\hat{b}_{j} = \frac{ R_{y} \mathcal{V}_{A} \bm{u}_{j} }{ \left\Vert R_{y} \mathcal{V}_{A} \bm{u}_{j} \right\Vert }.
	\end{equation}
	Moreover, plugging \eqref{c_j_beta_0} and \eqref{R_A_decomposition} into \eqref{Schur_n} shows that:
	\begin{equation}\label{schur}
	\mathcal{V}_{A}^{T} \mathcal{V}_{A} + i T_A \succeq \bm{c}_j \bm{c}_j^T = \mathcal{V}_{A}^{T} R_{y} \left( \hat{b}_{j} \hat{b}_{j}^{T} \right) R_{y} \mathcal{V}_{A}
	\end{equation}
	for all $ j \in \{ 1,...,n \} $.
	\begin{equation}\label{V_A_ineq}
	\mathcal{V}_{A}^{T} \left( I - R_{y} \left( \hat{b}_{j} \hat{b}_{j}^{T} \right) R_{y} \right) \mathcal{V}_{A} + i T_A \succeq 0
	\end{equation}
	Multiplying \eqref{V_A_ineq} by $ \bm{u}_{j}^{T} $ and $ \bm{u}_{j} $ on the left and right respectively, we have:
	\begin{equation}\label{key}
	\bm{u}_{j}^{T} \mathcal{V}_{A}^{T} \left( I - R_{y} \left( \hat{b}_{j} \hat{b}_{j}^{T} \right) R_{y} \right) \mathcal{V}_{A} \bm{u}_{j} \geq 0
	\end{equation}
	which is simply a rearrangement of \ref{Bob_inequality}, where \eqref{c_j_beta_0} and \eqref{T_A_irrelevant} (for $ \bm{u} = \bm{u}_j $) have been plugged in. This shows that the inequality \ref{Bob_inequality} is saturated if and only if $ \mathcal{V}_{A} \bm{u}_{j} $ belongs to the kernel of $ \left( I - R_{y} \left( \hat{b}_{j} \hat{b}_{j}^{T} \right) R_{y} \right) $. Since $ R_{y} \left( \hat{b}_{j} \hat{b}_{j}^{T} \right) R_{y} $ is a projection into $ span \left\{ R_{y} \hat{b}_{j} \right\} $, this occurs if and only if $ \mathcal{V}_{A} \bm{u}_{j} $ is parallel to $ R_{y} \hat{b}_{j} $, which indeed is satisfied by \eqref{Bob_optimal_choices}.
\end{proof}

\paragraph{\label{proof_harmonic}$\mathcal{P}_n$ is an harmonic function}
\begin{proof}
	First, let us denote: $ g_{i} \left( \vec{\mu} \right) \triangleq \sum_{j=1}^{i} \mu_{j} - i \cdot \mu_{i+1} $. Note that:
	\begin{equation}\label{exp_ln}
	P_{n} \left( \vec{\mu} \right) = \exp \left[ \ln \left( P_{n} \left( \vec{\mu} \right) \right) \right]
	\end{equation}
	Now, we find the second partial derivative of $ P_{n} $ relative to $ \mu_{k} $. \eqref{exp_ln} implies:
	\begin{equation}\label{key}
	\frac{ \partial^{2} }{ \partial \mu_{k}^{2} } P_{n} = \left( \frac{ \partial^{2} }{ \partial \mu_{k}^{2} } \ln P_{n} + \left[ \frac{\partial}{\partial \mu_{k}} \ln P_{n} \right]^{2} \right) \cdot P_{n}
	\end{equation}
	We shall show that $ \sum_{k=1}^{n} \frac{ \partial^{2} }{ \partial \mu_{k}^{2} } P_{n} \left( \vec{\mu} \right) = 0 $, by demonstrating that
	\begin{equation*}\label{key}
	\sum_{k=1}^{n} \left\{ \frac{ \partial^{2} }{ \partial \mu_{k}^{2} } \ln P_{n} + \left[ \frac{\partial}{\partial \mu_{k}} \ln P_{n} \right]^{2} \right\} = 0
	\end{equation*}
	In order to do so, we start by finding the first partial derivatives of $ \ln  P_{n} \left( \vec{\mu} \right) $.
	\begin{equation}\label{key}
	\frac{\partial}{\partial \mu_{k}} \ln P_{n} = \frac{1}{ \sum_{i=1}^{n} \mu_{i} } - \frac{k-1}{g_{k-1} } + \sum_{i=1}^{n-1} \frac{1}{g_{i} }
	\end{equation}
	and the second partial derivatives:
	\begin{equation}\label{key}
	\frac{\partial^{2}}{\partial \mu_{k}^{2}} \ln P_{n} = - \frac{1}{ \left( \sum_{i=1}^{n} \mu_{i} \right)^{2} } - \left( \frac{ k-1 }{g_{k-1} } \right)^{2} - \sum_{i=1}^{n-1} \frac{1}{ g_{i} ^{2} }
	\end{equation}
	thus,
\begin{eqnarray*}\label{key}
			&& \sum_{k=1}^{n} \left\{ \frac{ \partial^{2} }{ \partial \mu_{k}^{2} } \ln P_{n} + \left[ \frac{\partial}{\partial \mu_{k}} \ln P_{n} \right]^{2} \right\} =
			2 \sum_{k=1}^{n} \left\{ \frac{1}{\sum_{i=1}^{n} \mu_{i} } \left[ \sum_{i=k}^{n-1} \frac{1}{g_{i}  } - \frac{k-1}{g_{k-1} } \right] + \sum_{i=k}^{n-1} \frac{1}{g_{i} } \left[ \sum_{j=i+1}^{n-1} \frac{1}{g_{j} } - \frac{k-1}{g_{k-1}} \right] \right\} = \\ &&
			2 \left\{ \frac{1}{\sum_{i=1}^{k}} \left[ \sum_{i=1}^{n-1} \frac{i}{g_{i} } - \sum_{k=2}^{n} \frac{k-1}{g_{k-1} } \right] + \sum_{i=1}^{n-2} \sum_{j=i+1}^{n-1} \frac{i}{g_{i} } - \sum_{k=2}^{n} \frac{k-1}{g_{k-1} } \cdot \sum_{i=k}^{n-1} \frac{1}{g_{i}} \right\} = 0
		\end{eqnarray*}
which ends our proof.
\end{proof}

\end{widetext}

\end{document}